\def\Z_2{{Z \!\!\! Z}_2}
\def\[{[\![}
\def\]{]\!]}
\def\t{\theta}
\def\ra{\rangle}

\baselineskip 18pt

\hfill {\bf SISSA-93/96/FM}
\smallskip
\hfill {\bf IC/96/91}
\vskip 6mm

\noindent
{\bf An Analogue of Holstein-Primakoff and Dyson Realizations
for Lie Superalgebras. \hfill\break
The Lie superalgebra sl(1/n)}

\vskip 32pt
\noindent
T. D. Palev\footnote*{Permanent address: Institute for Nuclear
Research and Nuclear Energy, 1784 Sofia, Bulgaria; E-mail:
tpalev@inrne.acad.bg}

\noindent
International Centre for Theoretical Physics, P.O. Box 586,
34100 Trieste, Italy and  \hfill\break
International School for Advanced Studies, via Beirut 2-4,
34123 Trieste, Italy
\vskip 12pt

\vskip 48pt

{\bf Abstract.} An analogue of the Holstein-Primakoff and
of the Dyson realization for the Lie superalgebra $sl(1/n)$ is
written down. The expressions are the same as for the Lie
algebra $sl(n+1)$, however in the latter the Bose operators have
to be replaced with Fermi operators.

\vskip 48pt
In the present letter we extend the concept of 
Holstein-Primakoff (H-P) and of Dyson (D) realizations to 
case of Lie superalgebras and, more precisely, to the Lie
superalgebras $sl(1/n)$ for any $n$.

Initially the H-P and the D realizations were given for $sl(2)$
[1, 2]. It took some time the results to be generalized to the
Lie algebras $gl(n)$ [3], $so(2n)$ [4] and $sp(2n)$ [5]. The
extension to the case of quantum algebras is
available so far only for $sl(2)$ [6] and $sl(3)$ [7]. To our
best knowledge results on  H-P or D realizations for Lie
superalgebras were not published in the literature so far.

The motivation for the present work stems from the observation
that the Holstein-Primakoff and the Dyson realizations play
important role in various branches of physics. 
It is not possible to mention all contributions.
For applications in nuclear physics see [8, 9] and the references
therein; some early applications in the spin-wave theory
can be found in the book of Kittel [10] (see [11] for more recent
results), but there are, certainly, several other publications.

The H-P and D realization for $sl(1/n)$ we are going to present
is an analogue of the one given by Okubo [3] for the Lie algebra
$sl(n)$. Nevertheless it is only an analogue. The point is that
we give a realization of the generators in terms of Fermi
creation and annihilation operators (CAOs), whereas all H-P and D
realizations are imbeddings into enveloping algebras of Bose
CAOs. The similarity however is striking: replacing in the
expressions for $sl(n+1)$ the Bose CAOs with Fermi CAOs one
obtains the H-P and the D realizations for $sl(1/n)$.

We proceed to recall the definition of the Lie superalgebra (LS)
$sl(1/n)$ in the notation we are going to use. Let $U$ be the
(free complex) associative unital (= with unity) algebra of the
indeterminants $a_1^\pm,\ldots,a_n^\pm$ subject to the relations
(bellow and throughout $[x,y]=xy-yx$, $\{x,y\}=xy+yx$ )
$$
[\{a_{i}^+,a_{j}^-\},a_{k}^+]=\delta_{jk}a_{i}^+ -\delta_{ij}a_{k}^+,
\quad
[\{a_{i}^+,a_{j}^-\},a_{k}^-]=-\delta_{ik}a_{j}^- +\delta_{ij}a_{k}^-,
\quad
\{a_{i}^+,a_{j}^+\}=\{a_{i}^-,a_{j}^-\}=0.\eqno(1)
$$
For the sake of convenience we refer to $a_{i}^+$ and to
$a_{i}^-$ as to creation and annihilation operators,
respectively. Introduce a $\Z_2-$grading in $U$, setting
$$
deg(a_i^\pm)=\bar{1},\quad i=1,\ldots,n. \eqno(2)
$$
Then $U$ is an (infinite-dimensional) associative superalgebra,
which is also a LS with respect to the product (supercommutator)
defined between every two homogeneous elements $x, y \in U$ as
$$
\[x,y\]=xy-(-1)^{deg(x)deg(y)}yx; \eqno(3)
$$
its finite-dimensional subspace
$$
A={\rm lin.env.}\{a_i^\pm, \{a_j^+,a_k^- \} \vert
i,j,k=1,\ldots,n  \} 
\subset U \eqno(4) 
$$
is a Lie superalgebra with odd generators $a_1^\pm,\ldots,a_n^\pm$.

\noindent
PROPOSITION 1. {\it The {\rm LS}  $A$ is (isomorphic to) $sl(1/n)$,
$A=sl(1/n)$. $U$ is its universal enveloping algebra,
$U=U[sl(1/n)]$ }.

{\it Proof}. The proof can be found in [12]. We recall the main steps.
Let $e_{A,B}, \;\; A, B=0,1,2,\ldots,n$ be a $(n+1)\times (n+1)$
matrix with 1 on the intersection of the $A^{th}$ row and $B^{th}$
column and zero elsewhere. Then (in a lowest exact matrix
representation [13]) $sl(1/n)$ is a direct space sum of its
even subalgebra
$$
sl(1/n)_{\bar{0}}=
gl(n)=lin.env.\{e_{ij}+\delta_{ij}e_{00}|i,j=1,\ldots,n\} \eqno(5)
$$
and the odd subspace
$$
sl(1/n)_{\bar{1}}=lin.env.\{e_{0i},\; 
e_{i0}|i=1,\ldots,n \}.\eqno(6)
$$
The one to one correspondence between the representation
independent Cartan-Weyl generators (4) and the generators in the
above representation read:
$$
a_i^+ \leftrightarrow  e_{i0},\quad
a_i^- \leftrightarrow  e_{0i},\quad
\{a_i^+,a_j^-\} \leftrightarrow e_{ij}+\delta_{ij}e_{00},
\quad i,j=1,\ldots,n. \eqno(7)
$$
It is straightforward to check that the above map is a
LS isomorphism. In particular only from the relations (1) (and
the graded Jacoby identity) one derives also the commutation
relations between all even generators
$$
[\{a_i^+,a_j^-\},\{a_k^+,a_l^-\}]=\delta_{jk}\{a_i^+,a_l^-\}
-\delta_{il}\{a_k^+,a_j^-\}. \eqno(8)
$$
By definition $U[sl(1/n)]$ is the associative unital 
algebra of $a_1^\pm,\ldots,a_n^\pm$ subject to the
relations (1) and (8).  Since (8) follow from (1), one can skip
them. Therefore $U=U[sl(1/n)]$.

Let $f_i^\pm,\; i=1,\ldots,n$ be $n$ pairs of Fermi CAOs,
$$
\{f_i^+,f_j^- \}=\delta_{ij},\quad \{f_i^+,f_j^+ \}=
\{f_i^-,f_j^- \}=0.\eqno(9)   
$$
Denote by $W(0/n)$ the Fermi superalgebra, namely the
$4^{n}$-dimensional associative unital algebra with relations (9)
and $\Z_2$ grading induced from 
$deg(f_i^\pm)=\bar{1}$.

Consider the following operators in $W(0/n)$
$$
\varphi(a_i^+)=f_i^+,\quad \varphi(a_i^-)=(p-N)f_i^-, \quad
{\rm where}\;\;N=\sum_{k=1}^n f_k^+ f_k^-,\quad i=1,\ldots,n
.\eqno(10)
$$
It is an easy exercise to verify that the operators (10) satisfy
the relations (1) and moreover that they are linearly
independent. Thus, we have obtained the following result.

\noindent
PROPOSITION 2. {\it The map $\varphi$ defines  an isomorphism of
$sl(1/n)$ into $W(0/n)$ for any complex number $p$.}

This realization of $sl(1/n)$ in $W(n)$ is an analogue of the
Dyson realization of $sl(n+1)$ [2]. The latter can be formally
obtained form (10) replacing the Fermi operators with Bose CAOs
$$
[b_i^-,b_j^+]=\delta_{ij},\quad [b_i^+,b_j^+]=[b_i^-,b_j^-]=0.
\eqno(11)
$$
In that case the linear envelope of
$$
b_i^+,\quad (p-N)b_i^-, \quad 
{\rm and}\;\; [b_i^+, (p-N)b_j^-],\quad i,j=1,\ldots,n \eqno(12)
$$
span a realization of $gl(n+1)$.

The Dyson realization defines a $2^n-$dimensional representation of
$sl(1/n)$ in the Fermi Fock space $F(0/n)$. Let
$$
|p;\Theta\ra \equiv  |p;\t_{1},\t_{2},\ldots,\t_{n}\ra=
(f_{1}^+)^{\t_{1}}(f_{2}^+)^{\t_{2}}\ldots 
(f_{n}^+)^{\t_{n}}|0\ra,\quad \t_1,\ldots,\t_n=0,1,\eqno(13)
$$
be the usual orthonormed Fock basis in $F(0/n)$ and let
$|p;\Theta\ra_{\pm i}$ be a vector obtained from
$|p;\Theta\ra $ after the replacement of $\t_i$ with
$\t_i \pm 1$.
The transformations of the basis under the action of the
$sl(1/n)$ generators read:
$$
\varphi(a_i^+)|p;\Theta\ra=(1-\t_i)(-1)^{\t_1+\ldots+\t_{i-1}}
|p;\Theta\ra_{+i},\quad
\varphi(a_i^-)|p;\Theta\ra=\t_i(-1)^{\t_1+\ldots+\t_{i-1}}
(p-\sum_{k=1}^n \t_k +1)|p;\Theta\ra_{-i}.\eqno(14)
$$
In the generic case the representation of $sl(1/n)$ in
$F(0/n)$ is irreducible. If however $p=1,2,3,\ldots,n-1$,
the representation is indecomposible. Due to the factor 
$(p-\sum_{k=1}^n \t_k +1)$ the subspace of all vectors
with $\sum_{k=1}^n \t_k >p$ is an invariant subspace 
$F(0/n)_{inv}\subset F(0/n)$. The factor space
$F(0/n)_{irr}= F(0/n)/F(0/n)_{inv}$ is irreducible. By definition
such representations are called atypical [13].

The advantage of the Dyson realization (10) is its simplicity.
Its disadvantage stems from the observation that the
representation is not unitary, i.e., the hermitian conjugate to
$\varphi(a_i^-)$ is not equal to $\varphi(a_i^+)$, i.e.,
$$
(\varphi(a_i^-))^\dagger=\varphi(a_i^+)\eqno(15)
$$
does not hold.

It turns out all representations in $F(0/n)_{irr}$ corresponding
to integer nonnegative $p,\;\; p \in {\bf Z}_+$, are equivalent to unitary
representations. To this end one has to change in an appropriate
way the scalar product in $F(0/n)_{irr}$. We give the final results.

\noindent
PROPOSITION 3. {\it The map $\pi$ defined as
$$
\pi(a_i^+)=f_i^+\sqrt{p-N}, \quad \pi(a_i^-)=\sqrt{p-N}\;f_i^- \eqno(16)
$$
is an isomorphism of $sl(1/n)$ into $W(0/n)$ for any 
$p\in {\bf Z}_+$. The space $F(0/n)_{irr}$ with an orthonormed
basis consisting of all vectors
$$
|p;\Theta\ra \equiv  |p;\t_{1},\t_{2},\ldots,\t_{n}\ra,\quad
such \;\; that \quad \sum_{k=1}^n \t_k \le p,
\quad \t_1,\ldots,\t_n=0,1, \eqno(17)
$$ 
carries an unitary irreducible representation of $sl(1/n)$, i.e.,
(15) holds. The transformation of the basis reads:
$$
\eqalignno{
& \pi(a_i^+)|p;\Theta\ra=(1-\t_i)(-1)^{\t_1+\ldots+\t_{i-1}}
  \sqrt{p-\sum_{k=1}^n \t_k}\;|p;\Theta\ra_{+i},& (18a)\cr
& \pi(a_i^-)|p;\Theta\ra=\t_i(-1)^{\t_1+\ldots+\t_{i-1}}
  \sqrt{p-\sum_{k=1}^n \t_k +1}\;|p;\Theta\ra_{-i}.& (18b) \cr
}
$$
If $p<n$ the representation is atypical.}

The supercommutation relations (1) can be checked within the
particular representation, acting on the basis vectors (17)
according to (18).
One can proceed in a more algebraic way first expanding
$\sqrt{p-N}$ in a series with respect to $N$, so that it is
explicit that $\sqrt{p-N} \in W(0/n)$. 

The expressions (16) can be considered as analogue of the
Holstein-Primakoff realization for the Lie superalgebra $sl(1/n)$.
Also here, replacing the fermions in (16) with bosons one obtains
the H-P realizations for sl(n+1) as given by Okubo [3].

We wrote down explicit expressions in terms of fermions only for
the $\pi({a_i^\pm})$. The rest of the Cartan-Weyl generators,
namely $\pi( \{a_i^+,a_j^-\})=\{ \pi( a_i^+),\pi(a_j^-)\}$
follow from (16). One can express also the unit operator in
$F(0/n)_{irr}$ in terms of fermions, thus extending the results
to the LS gl(1/n). The more detailed results will be given
elsewhere directly for the LS gl(m/n).

Let us note that the explicit expressions for all
finite-dimensional irreducible representations of $sl(1/n)$ and
$gl(1/n)$ are known [14]. The formulas are however extremely
involved. Here we have obtained a small subset of all
representations, which description is however simple and is
realized in familiar for physics Fock spaces.

The above construction holds also for the case $n=\infty$. In
that case the representation space of $sl(1/\infty)$ has an
orthonormed basis, consisting of all vectors, such that
$\sum_{k=1}^\infty \t_k \le p$. All representations are 
atypical infinite-dimension irreducible representations.

One could expect that the above results may help to give H-P and
D realizations for the quantum algebra $U_q[sl(1/n)]$.  The
explicit representations of the Chevalley generators in a
Gelfand-Zetlin basis are known [15]. As a first step here one has
to pass to the corresponding deformed CAOs, expressing them
in terms of the Chevalley operators, and write subsequently the
transformation of the basis under the action of these operators.

\vskip 24pt
\noindent
{\bf Acknowledgments}

\bigskip
\noindent
The author is thankful to Prof. Randjbar-Daemi for the the kind
hospitality at the High Energy Section of ICTP, where most of the
results were obtained. He is grateful to Prof. C. Reina for making
it possible to visit the Section on Mathematical physics in
Sissa, where the paper was completed. The work was supported by
the Grant $\Phi-416$ of the Bulgarian Foundation for Scientific
Research.

\vskip 24pt
\noindent
{\bf References}

\vskip 12pt
\settabs \+  [11] & I. Patera, T. D. Palev, Theoretical 
   interpretation of the experiments on the elastic \cr 

\+ [1] & Holstein, T., and Primakoff, H., {\it Phys. Rev.} {\bf 58}, 
        1098 (1940). \cr            

\+ [2] & Dyson, F. J.,  {\it Phys. Rev.} {\bf 102}, 1217 (1956).\cr

\+ [3] & Okubo, S., {\it J. Math. Phys.} {\bf 16}, 528 (1975).\cr

\+ [4] & Papanicolaou, N., {\it Ann. Phys. (NY)} {\bf 136}, 210 
         (1981).\cr

\+ [5] & Deenen, J., and Quesne, C., {\it J. Math. Phys.} {\bf 23}, 
         878, 2004 (1982).\cr

\+ [6] & Chaichian, M., Ellinas, D., and Kulish, P. P., 
         {\it Phys. Rev. Lett.} {\bf 65}, 980 (1990). \cr
\+     & Quesne, C.,  {\it Phys. Lett. A} {\bf 153},
         303 (1991).\cr
\+     & Chakrabarti, R., and Jagannathan, R., {\it J. Phys. A} 
         {\bf 24},  L711 (1991).\cr
\+     & Katriel, J., Solomon, A. I.,  {\it J. Phys. A} {\bf 24}, 2093 
        (1991). \cr
\+     & Yu, Z. R., {\it J. Phys. A} {\bf 24}, L1321 (1991).\cr
\+     & Kundu, A., and Basu Mallich, B., {\it Phys. Lett. A} {\bf 156},
         175 (1991). \cr
\+     & Pan, F., {\it Chin. Phys. Lett.} {\bf 8}, 56 (1991).\cr

\+ [7] & da-Providencia, J., {\it J. Phys. A} {\bf 26}, 
         5845 (1993).\cr

\+ [8] & Klein, A., and Marshalek, E. R. {\it Rev. Mod. Phys.} 
         {\bf 63}, 375 (1991).\cr

\+ [9] & Ring, P., Schuck, P., {\it The Nuclear Many-Body Problem}, 
         Springer-Verlag, New York, Heidelberg, Berlin.\cr

\+ [10] & Kittel, C., {\it  Quantum Theory of Solids}, Willey, 
          New York (1963).\cr

\+ [11] & Caspers, W. J., {\it Spin Systems}, World Sci. Pub. Co., Inc.,
          Teanek, NJ (1989).\cr

\+ [12] & Palev, T. D., {\it J. Math. Phys.} {\bf 21}, 1293
          (1982).\cr 

\+ [13] & Kac, V. G., {\it Lect. Notes Math.} {\bf 676}, 597 (1978) \cr

\+ [14] & Palev, T. D., {\it J. Math. Phys.} {\bf 28}, 2280
          (1987); {\bf 29}, 2589 (1988), {\bf 30}, 1433 (1989).\cr 
\+       & Palev, T. D., {\it Funkt. Anal. Prilozh.} {\bf21}, \#3,
           85 (1987) (In Russian); \cr
\+		 & {\it Funct. Anal. Appl.}
		   {\bf 21}, 245 (1987) (English transl.)\cr

\+ [15] & Palev, T. D. and Tolstoy, V. N,, {\it Comm. Math. Phys.}
          {\bf 141}, 549 (1991). \cr
\end